\documentclass[conference, 10pt]{IEEEtran}
\usepackage[cmex10]{amsmath}

\usepackage{epsfig}
\usepackage{epstopdf}
\usepackage[tight]{subfigure}

\usepackage{algorithmic}
\usepackage{algorithm2e}
\usepackage{graphicx}
\usepackage{color}
\usepackage{float}
\usepackage{cite}

\begin{document}

\title{Sequential Sampling for Optimal Bayesian Classification of Sequencing Count Data}

\author{\IEEEauthorblockN{Ariana Broumand\textsuperscript{$\ast$} and Siamak Zamani Dadaneh \textsuperscript{$\ast$}}
\IEEEauthorblockA{\textsuperscript{$\ast$}Department of Electrical and Computer Engineering, Texas A\&M University,
  College Station, TX 77843, USA\\
  broumand@email.tamu.edu,  siamak@email.tamu.edu \\
}
}

\maketitle

\begin{abstract}
High throughput technologies have become the practice of choice for comparative studies in biomedical applications. Limited number of sample points due to sequencing cost or access to organisms of interest necessitates the development of efficient sample collections to maximize the power of downstream statistical analyses. We propose a method for sequentially choosing training samples under the Optimal Bayesian Classification framework. Specifically designed for RNA sequencing count data, the proposed method takes advantage of efficient Gibbs sampling procedure with closed-form updates. Our results shows enhanced classification accuracy, when compared to random sampling.
\end{abstract}


\section{Introduction}
\label{intro}
The advent of Next-Generation Sequencing technologies such as RNA sequencing (RNA-Seq), has provided a powerful tool to study transcriptome with unprecedented throughput, scalability, and speed \cite{Wang}.  In recent years, different distributions such as two stage poisson, hierarchical Gaussian-poisson and Negative Binomial (NB) have been used for modeling RNA-seq read counts \cite{Auer,Globalsip15,Robinson2007,Jason_hierarchy, Anders}. Due to over-dispersion property, NB distribution has gained the most attraction for modeling gene counts in RNA-Seq \cite{Globalsip15,Imani,siamak2,siamak1,siamak3}, and thereby we adopt it in this paper.

While modeling biological systems of any type, uncertainties show themselves in various forms. 
For instance while some studies have used data-driven methodologies to address dealing with uncertainties \cite{mokh1,mokh2,mokh3}, stochastic parametrization is often used to address uncertainties in computational models of brain sub-regions and brain connectivity\cite{R1,R2,R3}. Different algorithms are introduced for robust Bayesian network identification for pathway detection in biomedical data \cite{mandana1,mandana2,mandana3}. In \cite{rooz1,rooz2}, an experimental design framework for gene regulatory networks has been proposed, which quantifies the model uncertainty based on its induced intervention cost under a Bayesian framework.
In order to systematically quantify the model uncertainties
 and utilize the prior knowledge in addition to observed data, an optimal Bayesian classifer (OBC) has been introduced in \cite{OBC1,OBC2} which possesses minimum expected error across the uncertainty class governing the model. 
When doing classification (e.g. phenotypic classification) and regression (e.g. biomarker estimation) based on genomic data, where usually small samples are available and even labels might be missing,  incorporation of knowledge of pathways, regulating functions and other population statistics to construct prior distributions for optimal Bayesian classification and regression proves helpful \cite{shahin1,shahin2, arghavan}.

In contrast with the set of approaches that even in case of limited training data perform classification 
 based on a fixed available training set \cite{moti1,moti2,seyed}, we follow another direction and examine the optimal way to expand the training set in order to get the optimal Bayesian classifier under uncertainty \cite{Sequential}. 
  In case of being restricted to small sample data, random sampling is not optimal for classifier design \cite{Sequential, Zoll}. In this work, given a sample set $S_{n}$, consisting of $n$ data points, the goal is to select the next data point in such a way as to minimize the expected error of the optimal Bayesian classifier. Here with the assumption of NB distribution for prior class conditional probabilities, the challenge is the lack of a conjugate prior-posterior class conditional probability distribution. To overcome this issue, we have used MCMC approximation of the posteriors, to estimate classification error and choose the class that minimizes the expected error, to take the next sample point from. While this might sound conceptually similar to active learning or Bayesian experiment design paradigms \cite{sh3,sh4}, there are vital differences \cite{Sequential}. Those algorithms control the selection of potential unlabeled training points in the sample space to be labeled and used for further training. Similar to \cite{Sequential}, we generate new sample points from a chosen known label with direct target of reducing classification error. Reducing uncertainty in our class probability distributions is a secondary outcome. 

\section{Methods}
\label{methods}

\subsection{NB Model for RNA-Seq read counts in different classes}

We consider the expression level (count data) for $n$ genes in either class, sequenced at two conditions (classes) $0$ and $1$.  We write the distribution of $x_{gjk}$, the expression level of gene $g$ in sample $j$ for class $k$ as follows \cite{siamak1}:
\vspace{-.1cm}
\begin{equation}\label{nb}
	x_{gjk} \sim \text{NB}(r_g,p_{gk}),
\end{equation}

where $r_g>0$ is the dispersion parameter of NB distribution corresponding to gene $g$ and $0 < p_{gk} < 1$ is the probability parameter of NB distribution corresponding to gene $g$ and class $k$. The probability mass function (PMF) of $x_{gjk} \sim \text{NB}(r_g,p_{gk})$ is expressed as
\begin{equation}\label{formula}
 f_X(x_{gjk})=\frac{\Gamma(x_{gjk}+r_g)}{x_{gjk}!\Gamma(r_g)}p_{gk}^{x_{gjk}}(1-p_{gk})^{r_g}
 \end{equation} 
 where $\Gamma(\cdot)$ is the gamma function. 

We complete the model by placing conjugate priors on dispersion and probability parameters, in order to model the uncertainty we have about classes. More precisely, the prior distributions are as follows:

\begin{align}
	r_g  \sim & \mbox{Gamma}(e_0,\frac{1}{f_0}),\nonumber\\
	p_{gk}  \sim & \mbox{Beta}(a_0,b_0),
\end{align}

where $e_0$ and $f_0$ are shape and rate parameters of the gamma distribution, and $a_0$ and $b_0$ are hyperparameters of beta distribution.

\subsection{OBC and Error conditioned sampling}
Under the OBC framework, we assume that the actual model belongs to an \emph{uncertainty
class}, $\Theta $, of  feature-label distributions parameterized by $%
\boldsymbol{\theta }=[c, \boldsymbol{\theta }_{0}=\{R,P_0\},
\boldsymbol{\theta }_{1}=\{R,P_1\}]$
where $c=Pr(k=0)$ (the proportion of class 0 sample points in the population), $R$ is the vector of $r_g$ for all genes, shared for both classes, $P_0$ is the vector of $p_{g0}$ for all genes, in class $0$ and  $P_1$ is the vector of $p_{g1}$ for all genes, in class $1$   . 
Denoting a test point (in our case, a set of read counts for all genes) by $X$, in a 2 class binary classification framework
the optimal Bayesian classifier (OBC) is given by \cite{OBC1}
\begin{equation}
\psi _{obc}(X)=%
\begin{cases}
0 & \text{if } E_{\pi ^{\ast }}[c]f(X|0)\geq  (1-E_{\pi ^{\ast }}[c])f(X|1) \\ 
1 & \mathrm{otherwise},%
\end{cases}
\label{OBC_formula}
\end{equation}%
where 
\begin{equation}
f(X|k)=\int_{\Theta _{k}}f_{\boldsymbol{\theta }_{k}}(X|k)\pi ^{\ast }(%
\boldsymbol{\theta }_{k})d\boldsymbol{\theta }_{k},
\end{equation}%
is known as the \emph{effective density} and $E_{\pi ^{\ast }}[.]$ stands for the expectation over the posterior distribution.

We use the following terminology: \newline
$X_{n+1}:(n+1)^{th}$ sample point (random variable). A sample point consists of read counts for all the genes in the model.\newline
$\boldsymbol{u}^{n}:$ observed sample points from class 0 in the
existing $n$ sample points (observed).\newline
$\boldsymbol{v}^{n}:$ observed sample points from class 1 in the
existing $n$ sample points (observed). \newline
$\psi_{obc}^{n}:$ OBC designed with $\boldsymbol{u}^{n},\boldsymbol{v}^{n}, a_0, b_0, e_0, f_0$.\newline
$\tilde{\psi}_{obc}^{n+1}:$ OBC designed with $X_{n+1},\boldsymbol{u}^{n}, \boldsymbol{v}^{n}, a_0, b_0, e_0, f_0$.\newline
$\varepsilon ^{n+1}=$ Classification error of $\tilde{\psi}_{obc}^{n+1}$.

For a discrete classifier,
\begin{equation}
\varepsilon =c\underbrace{P(\psi (X)=1|k=0)}_{\varepsilon _{0}}+(1-c)%
\underbrace{P(\psi (X)=0|k=1)}_{\varepsilon _{1}}.  \label{eq:totError}
\end{equation}%
 
In our algorithm, summarized in Algorithm \ref{alg1}, we first compute the
expected cost (expected classification error) incurred if the next sample
point comes from each class:

\begin{equation}
\begin{array}{c}
e^{0}=E\Big[\varepsilon ^{n+1}|k_{n+1}=0,\mathbf{U}^n=\boldsymbol{u}^n,%
\mathbf{V}^n=\boldsymbol{v}^n\Big] \\ 
e^{1}=E\Big[\varepsilon ^{n+1}|k_{n+1}=1,\mathbf{U}^n=\boldsymbol{u}^n,%
\mathbf{V}^n=\boldsymbol{v}^n\Big].%
\end{array}
\label{ezero}
\end{equation}%
This notation simply means in the already observed training sample points, we have seen points ${u}^n$ from class $0$ and points ${v}^n$ form class $1$, and now we add a new training point once from class $0$ to have $\tilde{\psi}_{obc}^{n+1}$ that gives us $e^{0}$ and once from class $1$ to form $\tilde{\psi}_{obc}^{n+1}$ that gives us $e^{1}$.

\begin{algorithm}
\caption{Choosing which class to take the next sample point from}
\label{alg1}
\begin{algorithmic}[1]\small
\STATE{\textbf{input}: $c, {u}^n, {v}^n, a_0, b_0, e_0, f_0$}
\STATE{\textbf{output}: $k_{n+1}\in\{0,1\}$ : The class to take next sample point from}
\STATE{compute $e^{0}$ via Equation~\eqref{ezero}}
\STATE{compute $e^{1}$ via Equation~\eqref{ezero}}
	\IF {$e^{1}<e^{0}$} 
	\STATE {$k_{n+1}\gets 1$}
	\ELSE
	\STATE {$k_{n+1}\gets 0$}
	\ENDIF
	\STATE {Take (randomly) a new sample point from class $k_{n+1}$}
\end{algorithmic}
\end{algorithm}

\subsection{Inference via Gibbs sampling}
With the choice of model and prior distributions, there is no direct way to calculate \eqref{ezero}, but we approximate it using MCMC method.
In the following, we present our efficient MCMC inference of model parameters, which takes advantage of a novel data augmentation technique, leading to closed-form parameter updates.

The negative binomial random variable $n \sim \text{NB}(r,p)$ can be generated from a compound Poisson distribution as 
\begin{equation}
n = \sum_{t=1}^{\ell} u_t, \;\; u_t \sim \text{Log}(p), \;\; \ell \sim \text{Pois}(-r\ln (1-p)), \nonumber
\label{infe1}
\end{equation}

where $u \sim \text{Log}(p)$ corresponds to the logarithmic random variable \cite{johnson2005univariate}, with the pmf 
\begin{equation}
 f_U(u) = -\frac{p^u}{u\ln(1-p)}, u=1,2,...
\label{infe2}
\end{equation} 
As shown in \cite{zhou2015negative}, given $n$ and $r$, the distribution of $\ell$ is a Chinese Restaurant Table (CRT) distribution, 
\begin{equation}
(\ell | n,r) \sim \text{CRT}(n,r)
\label{infe3}
\end{equation}
 whose random samples  can be generated as 
 \begin{equation}
\ell =\sum_{t=1}^{n} u_t,~u_t  \sim 
\text{Bernoulli}(\frac{r}{r+t-1})
\label{infe4}
\end{equation}
.

Exploiting the above data augmentation technique, and in addition gamma-Poisson and beta-NB conjugacies, we can derive the update samplings of model parameters in a Gibbs sampling procedure as follows:

\begin{align}
	p_{gk}|- \sim & \mbox{Beta}(a_0 + \sum_{j} x_{gjk}, b_0 + n_k r_g),\nonumber\\
	\ell_{gjk} \sim & \mbox{CRT}(x_{gjk}, r_g),\nonumber\\
	r_g \sim & \mbox{Gamma}\Big(e_0 + \sum_{k,j} \ell_{gjk}, \frac{1}{f_0 - \sum_k n_k log(1-p_{gk})} \Big),
\end{align}

where $\cdot|-$ denotes conditioning on all the other parameters.

\begin{algorithm}
\caption{Computing Equation~\eqref{ezero}} 
Algorithm \ref{alg2} summarizes the use of MCMC method in calculating Eq. \eqref{ezero}.
\label{alg2}
\begin{algorithmic}[1]\small
\STATE{\textbf{input}: ${u}^n, {v}^n, a_0, b_0, e_0, f_0$}
\STATE{\textbf{output}: $e^{0}$ } 
\STATE{Update $R, P_0, P_1$ through Eq. \eqref{infe2},\eqref{infe3}, \eqref{infe4}.}
\STATE{Generate a set of test points $\{X_{test}\}$ from both classes according to $c, R, P_0, P_1$}
\STATE{$X_{n+1} \sim NB(R,P_0)$}
\STATE{Update $R, P_0, P_1$ with $X_{n+1}$, through Eq. \eqref{infe2}, \eqref{infe3}, \eqref{infe4}.}
\STATE{Form $\psi_{obc}^{n+1}$ using $c, R, P_0, P_1$}
\STATE{$e^{0} \gets$ mean( Classification error of  $\psi_{obc}^{n+1}$ on $\{X_{test}\}$ }
\STATE{return $e^{0}$}
\end{algorithmic}
\end{algorithm}

\section{Simulations and Results}
\label{results}
\begin{figure}[tbp]
\centering
{%
\subfigure[ $a_0=1$, $b_0=1$]{\includegraphics[scale=0.15]{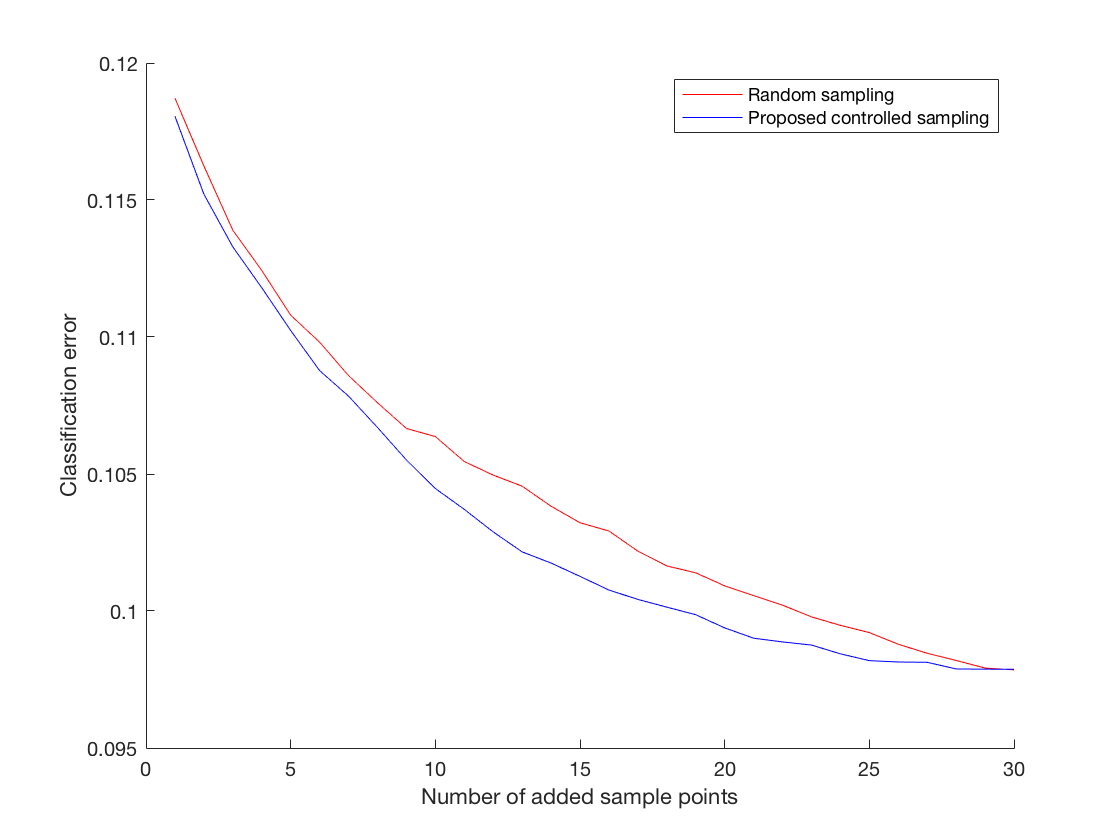}}%
} {%
\subfigure[$a_0=5$, $b_0=5$]{\includegraphics[scale=0.15]{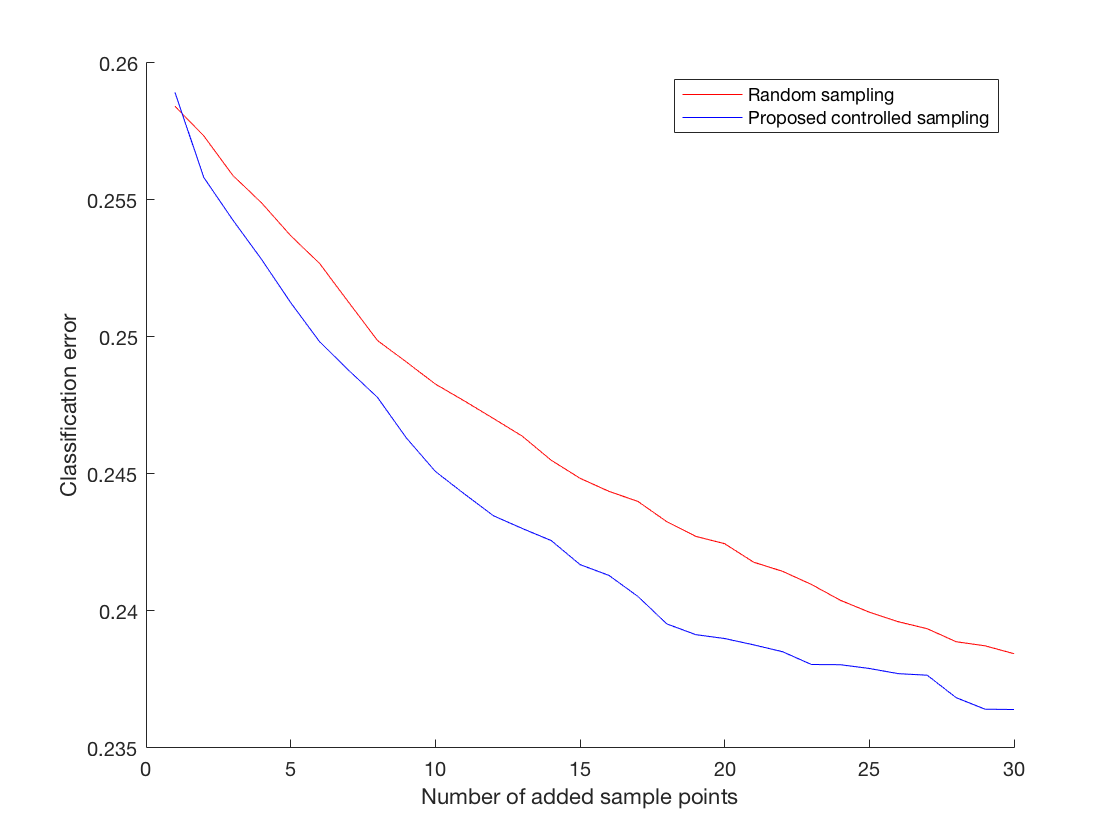}}%
} {%
\subfigure[$a_0=15$, $b_0=15$]{\includegraphics[scale=0.15]{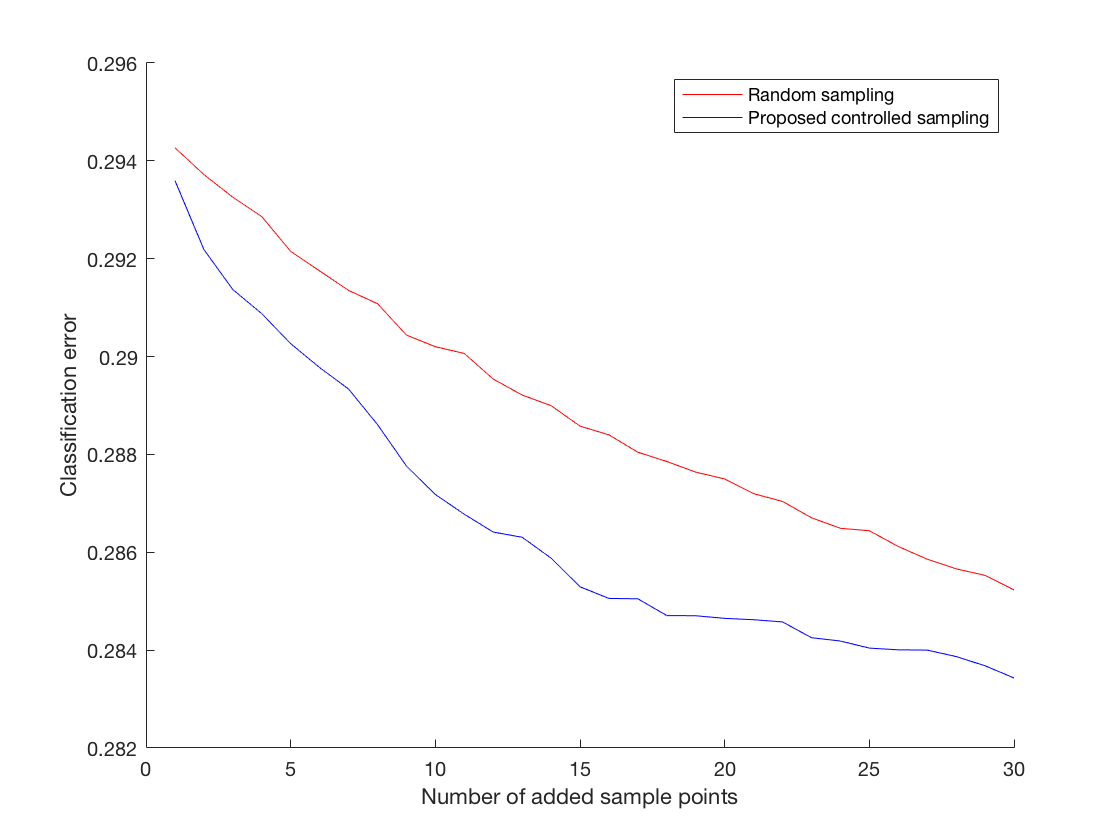}} }
\caption{{\protect\footnotesize Comparing performance of the proposed sequential sampling method with random sampling for $c=0.3$, $e_0=1$, $f_0=1$. In each simulation, average errors of optimal Bayesian classifiers at each training set size are reported for each sampling method (Red is random sampling, blue is the proposed method) .}}
\label{fig1}
\end{figure}

In order to assess the performance of the proposed sampling method, different simulations are run.  In each simulation, a fixed set of values is chosen for hyperparameters $c, a_0, b_0, e_0,f_0$. In different simulations, parameter values have been chosen in such a way to cover different levels of Bayes error and separability for the two class classification problem. Values of $c$ above $0.5$ are not considered, due to symmetry of the problem for two classes. 
\newline In each simulation, after fixing hyperparameters, a total of $10$ sample points are initially generated to populate class observations  as the initial state and a fixed set of $10,000$ test points with known labels is generated to evaluate classifier performances throughout the simulation. Each simulation consists of $2000$ repetition. In each repetition, 2 scenarios are compared. 
\newline In the first scenario, each time a new training point is randomly generated according to class proportions $c$ and $1-c$ and after addition of each new point, a new classifier is designed with all the available training points. This is continued until we obtain $30$ new training sample points. 
\newline In the second scenario, each time a new training point is generated from the class that is determined through our method in Section \ref{methods} and after addition of each new point, a new classifier is designed with all the available training points. This scenario too is continued until we obtain $30$ new training sample points. 
Average performance of classifiers given a certain size of training set is calculated for both methods, through averaging over $2000$ repetitions. 
Figure \ref{fig1} compares classification errors for the two methods, for $c=0.3$, $e_0=1$, $f_0=1$ and different values of $a_0 , b_0$. In our problem, two degrees of freedom regarding the Beta function can produce sufficient different levels of Bayes error to deem the study of simultaneous changes in all $4$ hyperparameters unnecessary. 

It can be seen that in all 3 graphs of Figure \ref{fig1}, the error curve for the proposed controlled sampling method lies below the curve of random sampling. This means that the training set obtained by the proposed method has helped develop an optimal Bayesian classifier with a lower average error than such a classifier trained on a randomly obtained (stratified according to $c$) training set, i.e. the proposed sampling method beats random sampling. It can further be noticed as the values of $a_0$ and $b_0$ increase, the classes become less separable, yielding higher values of classification error on the vertical axis. Variance of Beta distribution with parameters $a_0$ and $b_0$ is given by
\begin{equation}
\frac{a_0b_0}{(a_0+b_0)^2(a_0+b_0+1)}
\label{betavar}
\end{equation}
Hence higher values of $a_0$ and $b_0$ result in more similar classes. It can also be observed in Figure \ref{fig1} that with other parameters fixed, the efficacy of the proposed method is improved when we go above lower ranges of Bayes error around or below $0.1$.  Another observation which is expected is the that this gain in classification error is seen in small sample sizes as with big training sets enough information is already gathered to suppress the uncertainty in class conditional distributions, hence performances of both methods is expected to converge to each other, and in case of having infinitely many training points, to Bayes error of the problem. 

\begin{figure}[tbp]
\centering
{%
\subfigure[ $a_0=5$, $b_0=5$]{\includegraphics[scale=0.15]{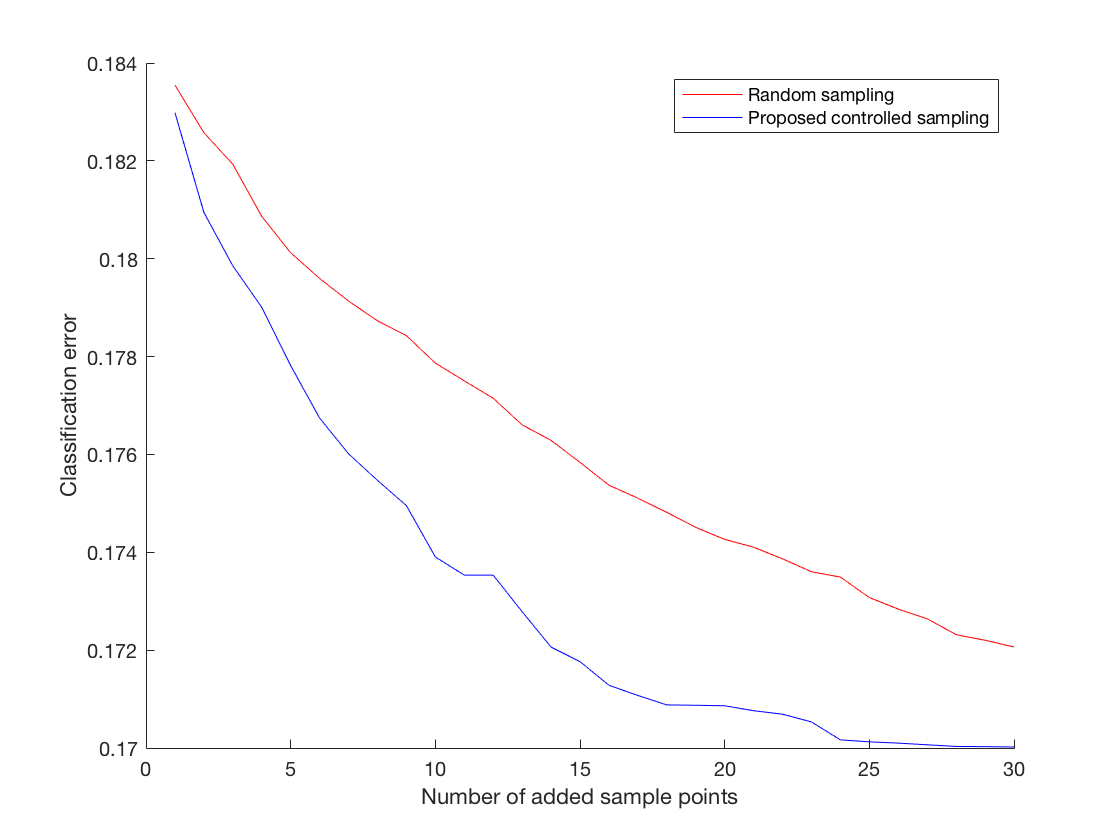}}%
} {%
\subfigure[$a_0=15$, $b_0=15$]{\includegraphics[scale=0.15]{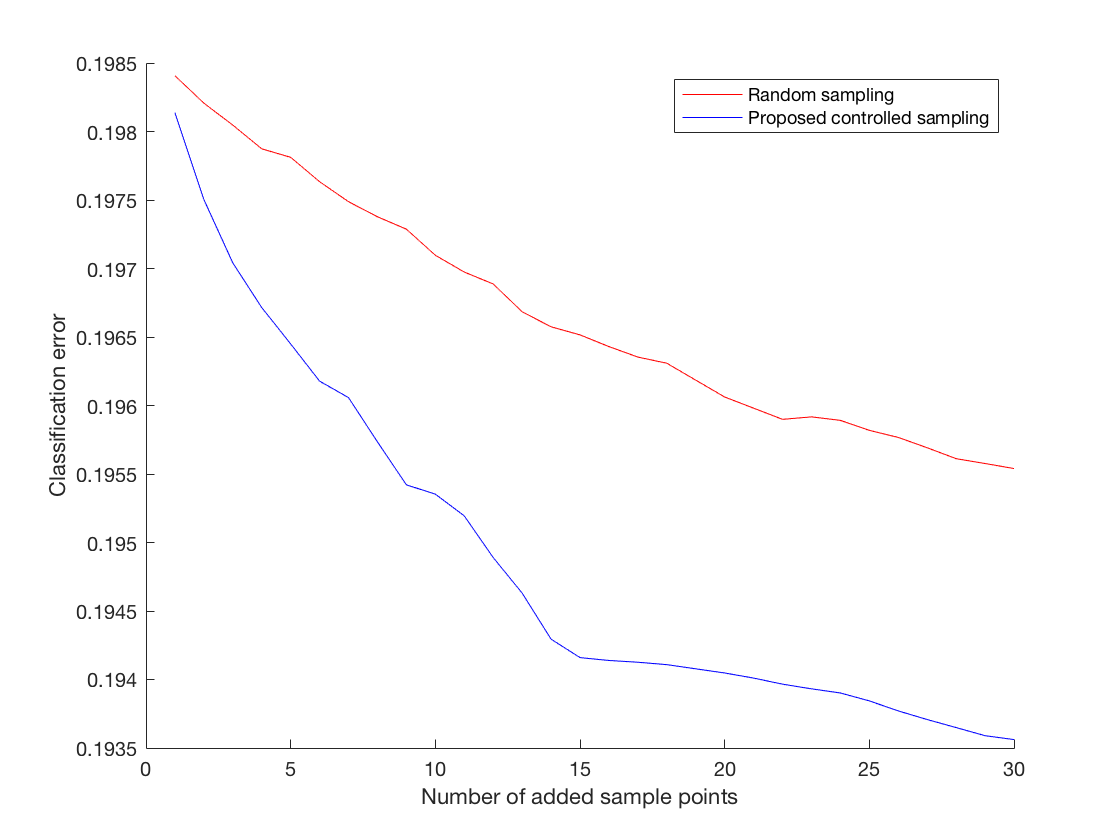}}%
} 
\caption{{\protect\footnotesize Comparing performance of the proposed sequential sampling method with random sampling for $c=0.2$, $e_0=1$, $f_0=1$. In each simulation, average errors of optimal Bayesian classifiers at each training set size are reported for each sampling method (Red is random sampling, blue is the proposed method) .}}
\label{fig2}
\end{figure}

Figure \ref{fig2} makes a similar comparison for the more unbalanced case of $c=0.2$. Proposed sampling method beats random sampling.
 
\begin{figure}[tbp]
\centering
{%
{\includegraphics[scale=0.15]{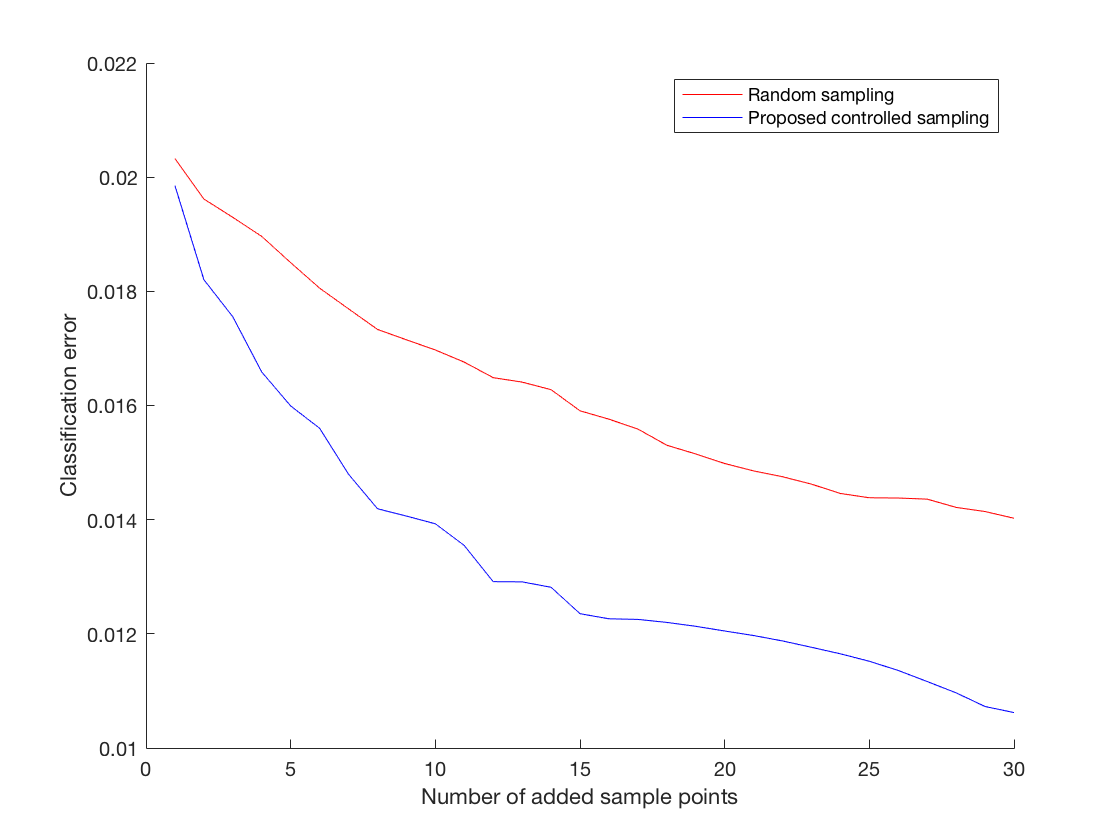}}%
}
\caption{{\protect\footnotesize Comparing performance of the proposed sequential sampling method with random sampling for $a_0=1$, $b_0=1$, $c=0.2$, $e_0=2$, $f_0=2$. In each simulation, average errors of optimal Bayesian classifiers at each training set size are reported for each sampling method (Red is random sampling, blue is the proposed method) .}}
\label{fig3}
\end{figure}

Figure \ref{fig3} shows a different case in the sense that values of $e_0$ and $f_0$ are increased to $2$ and $a_0$, $b_0$ are set equal to $1$ to produce a very low Bayes error. $c=0.2$. Proposed sampling method beats random sampling. It can also be seen that depending on the structure of the problem and parameter values, even in very low Bayes errors relatively high compared to other scenarios.

\begin{figure}[tbp]
\centering
{%
\subfigure[ $a_0=1$, $b_0=1$]{\includegraphics[scale=0.15]{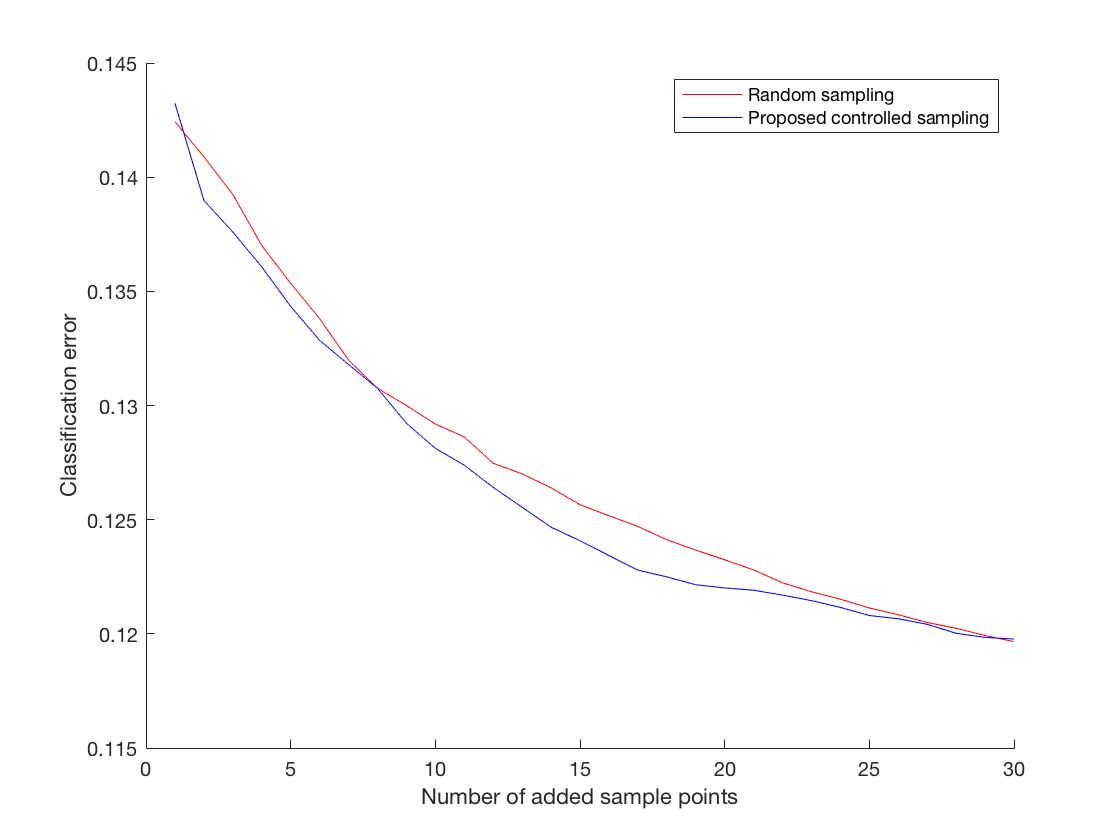}}%
} {%
\subfigure[$a_0=15$, $b_0=15$]{\includegraphics[scale=0.15]{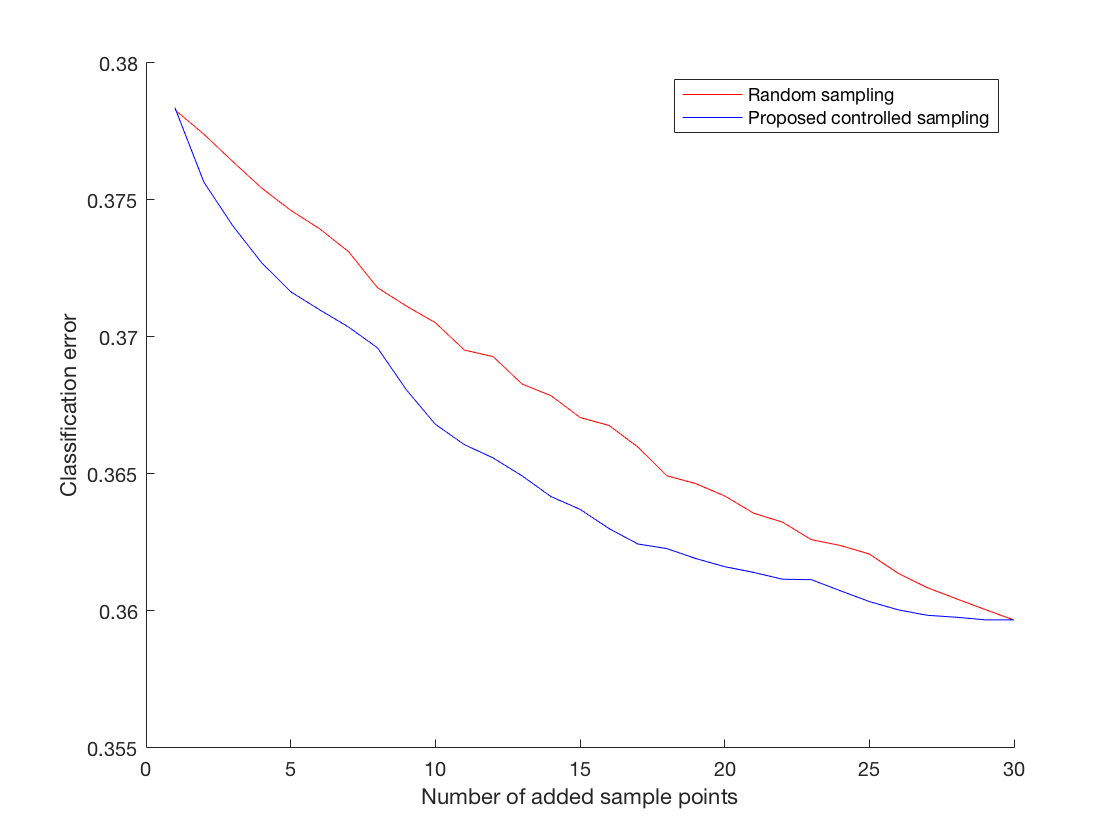}}%
} 
\caption{{\protect\footnotesize Comparing performance of the proposed sequential sampling method with random sampling for $c=0.4$, $e_0=1$, $f_0=1$. In each simulation, average errors of optimal Bayesian classifiers at each training set size are reported for each sampling method (Red is random sampling, blue is the proposed method) .}}
\label{fig4}
\end{figure}

Figure \ref{fig4} shows the simulation results for the more balanced class proportion of $c=0.4$. $e_0=1$ and $f_0=1$ have been used. While the proposed sampling algorithm still beats random sampling, it can be seen that the relative gain in classification error to the Bayes error is shrinking in this case. However this is not a big concern, as in most biological problems we deal with highly imbalanced data.

\section{Conclusions}
This study has further expanded the results of \cite{Sequential}, showing how to utilize the prior knowledge about classes to form training sets for classifiers more efficiently than random sampling, for negative binomial distributions which are used in modeling of RNA-Seq read counts. 
The method is described mathematically and its performance is theoretically studied using MCMC simulations on synthetic data. While efficiency of the proposed method seems to show a positive correlation with Bayes error, it is a function of all model parameters and hence can vary at the same levels of Bayes error.
Future works can include expansion of this sampling method to the hierarchical Gaussian-Poisson framework introduced in \cite{Jason_hierarchy} for modeling read count data, and also implementing the proposed sampling method on negative binomial models fit on real RNA-Seq datasets. Also employing convolutional neural networks as fully data-driven architectures to merge feature extraction and classification procedures like in \cite{Soleymani2018multi,Soleymani2018generalized} can be another area for potential research.
\newline
\newline
\newline

\end{document}